\documentclass[useAMS]{mn2e}
\usepackage{psfig}

\newcommand\arcdeg{\mbox{$^\circ$}}%

\title[He{\sc ii} Nebula in Holmberg II]{High Resolution Imaging of the
He{\sc ii} $\lambda$4686 Emission Line Nebula Associated with the
Ultraluminous X-Ray Source in Holmberg II}

\author[P. Kaaret et al.]{P.~Kaaret$^1$, M.J.~Ward$^2$, A.~Zezas$^1$ \\
$^1$Harvard-Smithsonian Center for Astrophysics, 60 Garden St.,
Cambridge, MA 02138, USA\\ $^2$Department of Physics and Astronomy,
University of Leicester, Leicester LE1 7RH, UK}

\date{Accepted . Received  ; in original form }

\pagerange{\pageref{firstpage}--\pageref{lastpage}} 
\pubyear{2004}

\begin{document}

\maketitle

\label{firstpage}

\begin{abstract}

We present Hubble Space Telescope images of the He{\sc iii} region
surrounding the bright X-ray source in the dwarf irregular galaxy
Holmberg II.  Using {\it Chandra}, we find a position for the X-ray
source of $\alpha=$ 08h 19m 28.98s, $\delta=$ +70\arcdeg 42\arcmin
19.\arcsec3 (J2000) with an uncertainty of 0.6$''$.  We identify a
bright, point-like optical counterpart centered in the nebula with the
X-ray source.  The optical magnitude and color of the counterpart are
consistent with a star with spectral type between O4V and B3 Ib at a
distance of 3.05~Mpc or reprocessed emission from an X-ray illuminated
accretion disk.  The nebular He{\sc ii} luminosity is $2.7 \times
10^{36} \rm \, erg \, s^{-1}$.  The morphology of the He{\sc ii},
H$\beta$, and [O{\sc i}] emission are consistent with being due to
X-ray photoionization and are inconsistent with narrow beaming of the
X-ray emission.  A spectral model consisting of a multicolor disk
blackbody with inverse-Compton emission from a hot corona gives a good
fit to X-ray spectra obtained with XMM-Newton.  Using the fitted X-ray
spectrum, we calculate the relation between the He{\sc ii} and X-ray
luminosity and find that the He{\sc ii} flux implies a lower bound on
the X-ray luminosity in the range 4 to $6 \times 10^{39} \rm \, erg \,
s^{-1}$ if the extrapolation of the X-ray spectrum between 54~eV and
300~eV is accurate.  A compact object mass of at least 25 to $40
M_{\odot}$ would be required to avoid violating the Eddington limit.

\end{abstract}

\begin{keywords} black hole physics -- galaxies: individual: Holmberg
II -- galaxies: starburst -- galaxies: stellar content -- X-rays:
galaxies \end{keywords}

\section{Introduction}

A key question in the study of the ultraluminous X-ray sources (ULXs)
is whether or not the X-ray emission is beamed.  If the X-rays are
beamed, then the ULXs may be accreting ``normal'' mass ($< 20
M_{\odot}$)  black holes or even neutron stars.  King et al.\ (2001)
have suggested that ULXs are high-mass X-ray binaries with
super-Eddington mass transfer rates in which the X-ray emission is 
funnelled, producing high observed X-ray fluxes for observers near the
beaming axis.  Motivated by the recent suggestion that jets may
contribute a significant fraction of the observed X-ray flux in
Galactic binaries previously thought to be dominated by disk and
coronal emission \cite{markoff01}, K\"ording et al.\ (2002) have
suggested that the ULXs may be stellar-mass black holes in which
relativistic beaming in jets aligned nearly along our line of sight
produce the high apparent X-ray fluxes.

If the ULXs are not strongly beamed, then the masses required for the
sources to be emitting below their Eddington luminosities are large,
well above the maximum possible mass ($\sim 20 M_{\sun}$) for a black
hole produced at the endpoint of the evolution of a normal star.  
Hence, the compact objects would be ``intermediate-mass'' black holes
\cite{colbert99}.  This has potentially interesting consequences.  
Intermediate-mass black holes may be excellent sources of gravitational
radiation \cite{ebisuzaki01,miller02}. They may be relics of the first
generation of star formation \cite{madau01}, where, due to the absence
of metals, extremely massive  stars were likely to have formed
\cite{larson98}. Or, they may be important in the formation of
supermassive black holes \cite{ptak99}.

Pakull \& Mironi (2002) discovered an He{\sc ii} $\lambda 4686$
emission line nebula in the ROSAT error box for the ULX in the dwarf
irregular galaxy Holmberg II at a distance of 3.05~Mpc
\cite{hoessel98}.  X-ray emission was first detected from Holmberg II
in the ROSAT all-sky survey and then localized in ROSAT HRI
observations to a single, highly variable source with a maximum
luminosity (if isotropic) of $\sim 10^{40} \rm \, erg \, s^{-1}$
\cite{zezas99}.  If a ULX is embedded in a diffuse nebula, then the
X-radiation of the ULX should photoionize the nebula, as seen for the
nebula surrounding LMC X-1 \cite{pakull86}.  Photoionization produces
(at most) one optical/UV photon for each X-ray photon doing the
excitation, so the line flux from the nebula is a direct measure of the
total number of ionizing photons emitted in all directions.  For
example, measurement of the He{\sc ii} $\lambda 4686$ flux gives a
direct count of all source photons with energies above 54~eV which
ionize the nebula.  The discovery of an He{\sc iii} region near a ULX
offers the exciting prospect of determining observationally whether or
not the X-ray emission from the ULX is beamed.  If the ULX is truly
ultraluminous, then the X-ray photoionization should produce observable
UV and optical line emission from high excitation levels.  Pakull \&
Mironi (2002) report an inferred photoionization luminosity of $(3 -
13) \times 10^{39} \rm \, erg \, s^{-1}$ for a distance of 3.2~Mpc
based on an assumed thermal bremsstrahlung X-ray spectrum, X-ray
spectral measurements from the ROSAT PSPC \cite{fourniol98} and ASCA
\cite{miyaji01}, and modelling of the photoionization nebula.  The
lower end of their allowed luminosity range is consistent with the
highest luminosities observed from stellar-mass black hole candidate
X-ray binaries within the Milky Way.

Here, we report on Hubble Space Telescope observations of the vicinity
of the ULX in Holmberg II taken in the optical emission lines He{\sc
ii} $\lambda 4686$, H$\beta$, and [O{\sc i}] $\lambda 6300$, new
Chandra High-Resolution Camera observations, and archival XMM-Newton
data.  The improved X-ray position from Chandra strengthens the
association of the nebula with the X-ray source.  The HST images enable
detailed inspection of the morphology of the nebula and allow us to
isolate it from other nearby nebulosities.  The XMM-Newton data provide
high quality information on the X-ray spectrum which are essential to
constraining the true luminosity of the system.  We describe the
observations and analysis in \S~2, and discuss the implications in
\S~3.

\section{Observations and Analysis}

\subsection{Chandra Observations}

Observations of Holmberg II were made using the High-Resolution Camera
(HRC) onboard the Chandra X-Ray Observatory (ObsID 3816; PI Kaaret).  
The HRC was used and the source was placed off-axis to prevent pile-up
from affecting the position determination.  The {\it Chandra\/}
observation began on 3 July 2003 05:56:20 UT and had a useful exposure
of 4.9~ks.  Chandra data were subjected to standard processing and
event screening.  No strong background flares were found, so the entire
observation was used.  A bright source is detected at high significance
which we identify with the ULX in Holmberg II \cite{zezas99,kerp02}. 
The {\it Chandra} position is: $\alpha=$ 08h 19m 28.98s, $\delta=$
+70\arcdeg 42\arcmin 19.\arcsec3 (J2000).  This position is determined
using the {\it Chandra} aspect solution, which has an estimated
uncertainty of 0.6$''$ at 90\% confidence (see the Chandra aspect pages
at http://cxc.harvard.edu/cal/ASPECT/celmon).  The X-ray image is
consistent with that expected for a point source at the given position
off-axis.  We find no evidence for a significant extended component as
reported by Miyaji et al.\ (2001).

\begin{table*}
\begin{center}
\begin{tabular}{lccccc}
\hline
Observation & $kT_{\rm in}$        & $kT_e$      &  $\tau$             & $L_{\rm X}$              & $L_{\rm PI}$          \\
          & [keV]                  & [keV]       &                     & [erg cm$^{-2}$ s$^{-1}$] & [erg s$^{-1}$] \\ \hline
10 April  & $0.22^{+0.02}_{-0.08}$ & $49 \pm 14$ & $1.1^{+0.3}_{-0.4}$ & $16 \times 10^{39}$      & $5.9 \times 10^{39}$ \\
16 April  & $0.20 \pm 0.02       $ & $76 \pm 23$ & $0.8 \pm 0.3 $      & $17 \times 10^{39}$      & $6.1 \times 10^{39}$ \\
19 Sept   & $0.17 \pm 0.02       $ & $71 \pm 30$ & $0.4^{+0.4}_{-0.2}$ & $ 5 \times 10^{39}$      & $3.7 \times 10^{39}$ \\ \hline
\end{tabular}
\end{center}

\caption{X-Ray Spectral Fits.  Best fit spectral parameters for each
observation.  The table includes: the date in 2002 of the observation;
the temperature of the multicolor disk blackbody emission, $kT_{\rm
in}$; the temperature, $kT_e$, and optical depth, $\tau$, of the hot
electron corona; the total intrinsic X-ray luminosity inferred from the
X-ray flux and the Comptonization model parameters assuming isotropic
emission, $L_{\rm X}$; and the total photoionization luminosity,
$L_{\rm PI}$, required to produce the observed He{\sc ii} luminosity
using the fitted X-ray spectrum as input to {\it Cloudy}.  The errors
correspond to 90\% confidence for a single parameter ($\delta chi^2 =
2.71$).} \label{specfits}

\end{table*}

\subsection{XMM-Newton data}

We extracted observations of Holmberg II from the XMM-Newton archive.
Analysis of these data have been previously reported by Dewangan et
al.\ (2004).  We reduced the data and generated response matrices using
the SAS software version 5.4.1.  We use only data from the EPIC-PN.  We
examined the three useful available observations.  The first
observation occurred on 10 April 2002 (ID 0112520601; PI Watson).  The
time on-target was  5.1~ks, during which the total PN rate was always
below 20~c/s indicating there were no background flares.  The total
exposure after live time correction was 4.7~ks.  The second observation
occurred on 16 April 2002 (ID 0112520701; PI Watson).  There were
several background flares, so we filtered the data to remove times with
PN count rates above 35~c/s.  The third observation occurred on 19 Sept
2002 (ID 0112520901; PI Watson). There were, again, background flares
and we filtered the data to removed times with PN count rates above
20~c/s.  Spectra were extracted from circular regions with a radius of
$24\arcsec$.   For all observations, the source was located near a chip
edge, and we used circular background regions on the same chip.  The
spectra were regrouped to have at least 100 counts per bin for the
first two observations and 25 counts per bin for the last.  Models were
fitted to the spectra using XSPEC version 11.2 over an energy band from
0.3 to 8~keV.

As found by Dewangan et al.\ (2004), complex or multicomponent models
are required to obtain adequate fits.  A model consisting of the sum of
powerlaw plus disk blackbody emission each modified by interstellar
absorption has often been used to model the spectra of black hole X-ray
binaries and ultraluminous X-ray sources.  Using such a model, we find
best-fitting parameters similar to those reported by Dewangan et al.\
(2004).

A key input to the photoionization modelling is the X-ray spectrum.  To
input to the photoionization code, we must extrapolate the fitted X-ray
spectrum to energies below the minimum observed X-ray energy; the shape
of the spectrum between 54~eV and 300~eV is important in determining
the relation between the He{\sc ii} luminosity and the X-ray
luminosity.  In the powerlaw plus disk blackbody model, the powerlaw
component is an adhoc model for a Compton emission component arising
from disk photons inverse-Compton scattered by energetic electrons in a
corona.  This suggests that the powerlaw component should be cut off at
energies below the disk temperature.  However, the model contains no
prescription for the form of such a cutoff.  We require a model in
which the low energy extension of the spectrum is well defined and
physically motivated.  Based on the application of Comptonization
models for detailed physical modelling of the multiwavelength spectral
of black hole X-ray binaries, we choose to use a Comptonization model
\cite{poutanen96}.   Specifically, we use the model of Poutanen \&
Svensson (1996), {\it compps} in XSPEC, to calculate the spectrum
produced by thermal Comptonization of photons emitted with a disk
blackbody spectrum and scattered in a hot corona. The model parameters
allowed to vary in the fits are the disk temperature $kT_{\rm in}$, the
corona electron temperature $kT_{e}$, and the corona optical depth
$\tau$.

Because the metallicity of Holmberg II is significantly lower than
solar, we use two absorption components: one to model absorption within
the Milky Way for which we fix the metallicity at solar and fix the
absorption column density $N_{\rm H} = (3.42 \pm 0.3) \times 10^{20}
\rm \, cm^{-2}$, and a second to model absorption within Holmberg II
for which we fix the  metallicity at $Z = 0.07 Z_{\sun}$
\cite{mirioni02}.  We performed a simultaneous fit to the data for all
three observations in which the absorption within Holmberg II was the
same for all observations and the Comptonization model parameters were
allowed to vary individually for each observation.  We found an
adequate fit with $\chi^2$/DoF = 273.5/285.  The best fit parameters
are reported in Table~\ref{specfits}.  The best fit column density is
$(3.7 \pm 0.5) \times 10^{21} \rm \, cm^{-2}$.  We note that this
column density is significantly above that found by Dewangan et al.\
(2004) because the absorption, beyond the Galactic component, is
calculated for the low metallicity appropriate to Holmberg II.

The three XMM-Newton observations cover the extremes of X-ray flux
detected from Holmberg II X-1 \cite{dewangan04}.  Therefore, these
observations also likely cover the range of X-ray spectral variations
in the source.  

We note a thermal bremsstrahlung model as used by Pakull \& Mirioni
(2002) for input to their photoionization modelling does not provide an
adequate fit to any of the observations, even with use of two distinct
absorption components.

\begin{figure*}
\centerline{\psfig{file=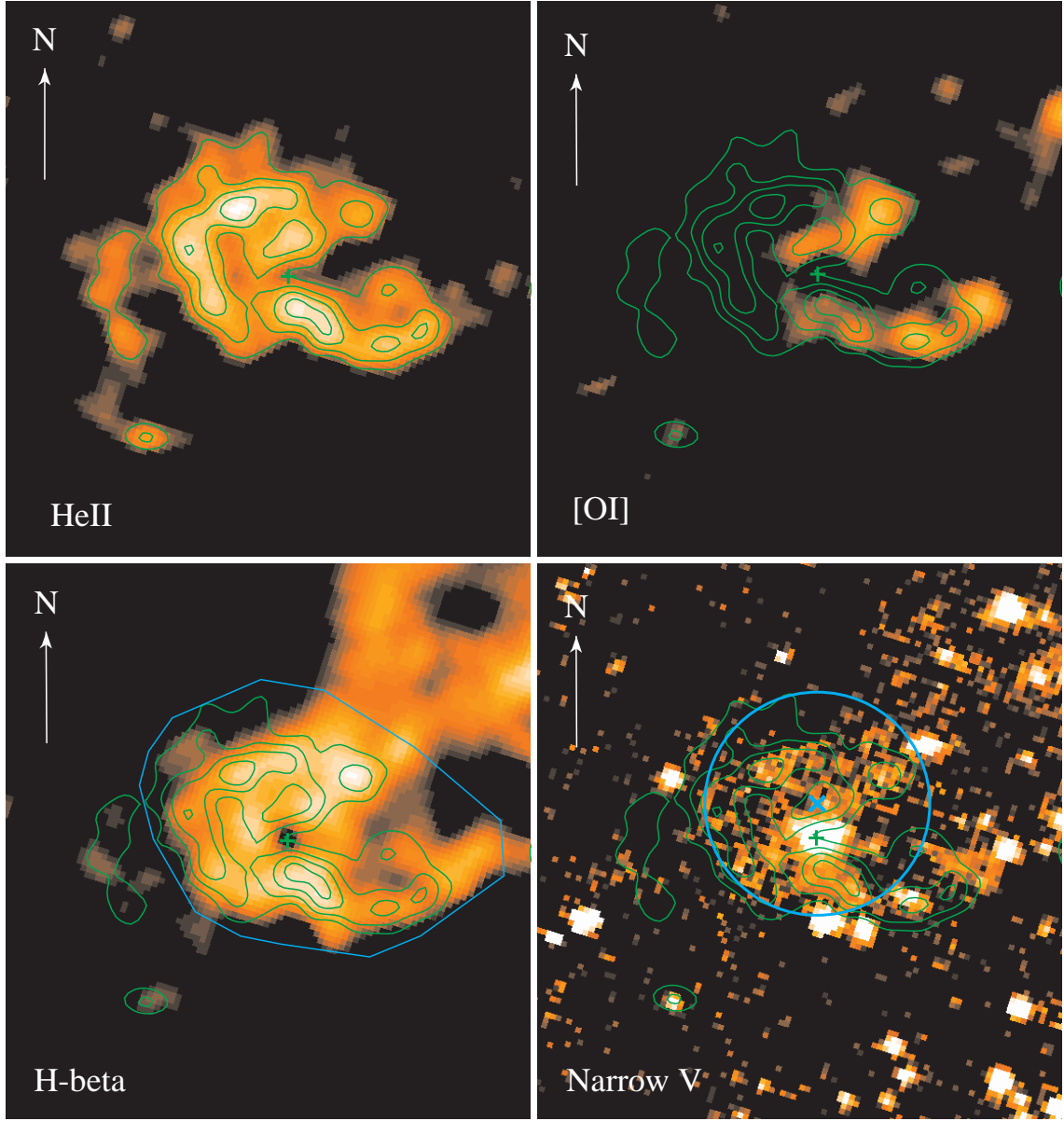,width=6.0in}}  \caption{False
color images of the continuum subtracted line emission for He{\sc ii}
$\lambda$4686, H$\beta$ $\lambda$4861, and [O{\sc i}] $\lambda$6300 and
for the narrow V-band continuum.  In each image, the arrow points
Northward and has a length of $1\arcsec$ (15~pc), East is to the left,
the green cross marks the position of the bright star within the
nebula, and the green curves are contours of the He{\sc ii} emission. 
The contour levels  are $[2,3,4,5] \times 1.9 \times 10^{-16} \rm \,
erg \, cm^{-2} \, s^{-1} \, arcsec^{-2}$.  The same contours are
plotted on all four panels.  The cyan circle and `X' in the narrow-V
image denote the best Chandra position for the ULX and the relative
Chandra/HST error circle.  The cyan polygon in the H$\beta$ image shows
the region used for the line flux measurements described in the text. 
Each line image was smoothed with a gaussian filter with a FWHM of
$0.24\arcsec$.} \label{hstimages} \end{figure*}

\subsection{HST observations}

Observations were made of Holmberg II centered on the position of the
ULX using the Advanced Camera for Surveys (ACS) on the Hubble Space
Telescope (HST) under GO program 9684 (PI Kaaret). Observations were
made in the narrow band filters FR462N centered on He{\sc ii}
$\lambda$4686, FR505N centered on H$\beta$ $\lambda$4861, and FR656N
centered on [O{\sc i}] $\lambda$6300.  Images in the FR462N filter 
were obtained on 21 January 2003 and 25 June 2003.  For the January
observation, the aimpoint was placed closed to the ramp filter edge and
there are non-uniformities in transmission across the image.  For this
reason, we quote fluxes for the He{\sc ii} emission based only on the
June observation.  The images in all of the other filters were obtained
on 24 November 2002.  All of the narrow band filters have a 2\%
bandwidth.  The recession velocity of Holmberg II is 157~km/s
\cite{strauss92}, so the redshifted emission lines lie within the
filter bandpasses.  In addition, observations were made in the medium
filters FR459M and F550M for continuum imaging and continuum
subtraction.

The standard processing for ACS data does not perform cosmic-ray
removal for images without cosmic-ray splits.  Because all of our
observations except those in the FR462N filter were performed in dither
patterns without cosmic-ray splits, we re-processed all observations
using the {\it PyRAF} task {\it multidrizzle} in STSDAS 3.1 to remove
the cosmic ray hits.  We found residual sky level offsets in the images
and removed these by fitting a gaussian to those pixels not containing
astronomical objects in a $30\arcsec \times 30\arcsec$ field centered
near the ULX and subtracting off the gaussian centroid.  We aligned the
F550M (narrow V band) image to 10 stars selected from the USNO A2.0
catalog \cite{monet96} using the {\it imwcs} tool from the Smithsonian
Astrophysical Observatory Telescope Data Center.  Based on the residual
offsets for the 10 stars, we estimate that the astrometric uncertainty
is $0.3\arcsec$.  We then aligned each other image to the aspect
corrected F550M image using the {\it IRAF} tools {\it geomap} and {\it
geotran} \cite{tody93}.  We checked the alignment using the tool {\it
xregister} and found that the residual misalignments were less than
0.1~pixel.  

We produced continuum subtracted images using the FR459M image to
estimate the continuum for the FR462N He{\sc ii} $\lambda$4686 and
FR505N H$\beta$ $\lambda$4861 images, and the F550M image to estimate
the continuum for the FR656N [O{\sc i}] $\lambda$6300 image.  Since we
are interested primarily in the diffuse, nebular emission, we located
the stars in each frame and subtracted off the stellar emission before
performing the continuum subtraction.  We fit a Moffat profile to
several bright stars to determine the point spread function shape, and
then used that fixed shape in fitting and subtracting the stars. We
note that the FR459M image used for continuum subtraction of the FR462N
image contains the He{\sc ii} line.  We correct for the apparent
reduction in the He{\sc ii} line flux caused by partial subtraction of
the line as described below.

For the continuum subtraction of the nebula, we assumed that the
intrinsic continuum spectrum is flat, $F(\lambda) \propto \lambda^{n}$
with $n = 0$ and reddened with an extinction of E(B-V) = 0.024
\cite{stewart00}.  For the He{\sc ii} and H$\beta$ images, the
continuum band lies close to the line wavelength, and changing the
continuum slope does not significantly affect the estimated line flux
(we discuss this further below in the estimation of the total He{\sc
ii} line flux).  For [O{\sc i}], the line wavelength lies further from
the best continuum band available.  For [O{\sc i}], we repeated this
procedure using continuum slopes $n = +2$ and $n = -2$. We found that
the morphology of the [O{\sc i}] emission varied little.
Fig.~\ref{hstimages} shows the continuum subtracted images in the three
lines and the F550M (narrow V band) continuum image.  The He{\sc iii}
nebula reported by Pakull and Mironi (2002) is clearly detected and is
at a position consistent with that of the X-ray source.  

There is one bright point source within the nebula visible in the
narrow V-band image.  The source profile appears similar to those of
other stars in each image and there is no evidence for spatial extent. 
The source appears to be a star, but a compact core of the nebula,
unresolved at the resolution of the ACS/WFC, could also contribute to
the emission.  The position of the star is $\alpha=$ 08h 19m 28s.99 and
$\delta=$ +70$\arcdeg$ 42$\arcmin$ 19$\arcsec$.0 (J2000).  This is
$0.3\arcsec$ from the Chandra position for the X-ray source.  This
offset is less than the relative astrometric uncertainty between the
Chandra and HST images.  Therefore, we consider the optical star to be
the counterpart of the X-ray source.  We find the ST magnitudes of the
star to be $22.04 \pm 0.08$ in the F550M filter and $21.35 \pm 0.10$ in
the FR459M filter.  The dominant uncertainty is in the removal of the
underlying nebular component.  If we consider only the extinction
within the Milky Way, then the reddening-corrected equivalent Johnson V
magnitude is $21.86 \pm 0.09$ and the color $({\rm B} - {\rm V})_{0} =
-0.20 \pm 0.15$.  Conversion to standard Johnson magnitudes increases
the quoted uncertainty because the filters used only approximately
match the Johnson bands.  There may also be extinction within Holmberg
II.   Using the relation between $N_{\rm H}$ and $E(B-V)$ determined
for the SMC \cite{bouchet85}, we find $E(B-V) = 0.07 \pm 0.01$ based on
our spectral fit to XMM-Newton data.  This is an upper bound on the
reddening since it is not clear whether the column density observed
towards the X-ray source necessarily obscures the companion star.  In
some Galactic black holes, the X-ray absorption is seen to vary on a
time scale of minutes, indicating that the absorption occurs close to
the compact object \cite{tomsick98}. Applying this additional reddening
with an SMC extinction law, we find a Johnson V magnitude of $21.64 \pm
0.11$ and a color $({\rm B} - {\rm V})_{0} = -0.25 \pm 0.16$.

To estimate the He{\sc ii} $\lambda 4686$ line flux from the nebula, we
found the total count rates for the nebula in the FR462N and FR459M
images from which the stellar emission had been subtracted.  We defined
the nebula using a contour drawn to enclose the He{\sc ii} and H$\beta$
emission which is shown as a cyan curve in the H$\beta$ image in
Fig.~\ref{hstimages}.  The FR459M filter bandpass included the He{\sc
ii} $\lambda 4686$ emission line (unfortunately, the FR459M bandpass is
too wide to exclude both He{\sc ii} and the stronger H$\beta$ and
H$\gamma$ emission from the nebula). Hence, we must solve for the
He{\sc ii} line flux and the continuum flux given the nebular count
rates in the two filters.  We assumed that the nebular continuum
emission has a powerlaw spectrum $F(\lambda) \propto \lambda^{n}$.  We
used the {\it IRAF} tool {\it calcphot} in the {\it synphot} package
with HST/ACS calibration files released on 10 Dec 2003 to find count
rates in each filter for given continuum flux and He{\sc ii} line
flux.  We then inverted the matrix relating the fluxes to count rates
and calculated the fluxes from the measured count rates.  For a flat
continuum with $n = 0$, we find that the flux in the He{\sc ii} line is
$2.4 \times 10^{-15} \rm \, erg \, cm^{-2} \, s^{-1}$.  Our continuum
subtraction procedure depends on the assumed continuum shape. 
Repeating the procedure for $n = +2$ and $n = -2$, we find that the
uncertainty in the He{\sc ii} line flux induced by the uncertainty in
the continuum slope is of order 3\%.  Including additional factors such
as the uncertainties in the subtraction of the stars, the calibrations
of the filters, and the subtraction of the sky background, we estimate
an overall uncertainty in our He{\sc ii} flux measurement of 9\%.  

Using the same contour defined for the He{\sc ii}, which was drawn to
enclose both the He{\sc ii} and H$\beta$ emission, we found the flux in
H$\beta$ of $1.1 \times 10^{-14} \rm \, erg \, cm^{-2} \, s^{-1}$ and
in [O{\sc i}] of $1.3 \times 10^{-15} \rm \, erg \, cm^{-2} \, s^{-1}$.
The uncertainty on the H$\beta$ flux is 5\% and on the [O{\sc i}]  is
20\%.  We find a ratio He{\sc ii}/H$\beta = 0.22 \pm 0.02$.  This is
larger than the value of 0.16 found by Pakull and Mironi (2002), but
the difference may be due to sampling of a larger spatial region due to
the effect of seeing in their ground-based spectrum.  To test this, we
convolved our image with a gaussian with $2.0\arcsec$ FWHM and then
extracted fluxes from a circular region with the same area as the
contour defined in Fig.~\ref{hstimages}.  We find He{\sc ii}/H$\beta =
0.18 \pm 0.02$, consistent with Pakull and Mironi (2002).  The ratio is
at the extreme end of the range found in photoionized nebulae in nearby
galaxies \cite{garnett91}.


\begin{figure} \centerline{\psfig{file=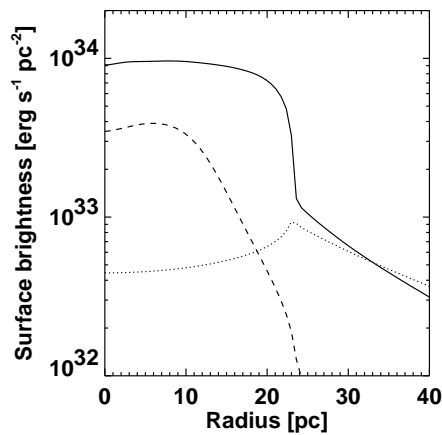,width=2.5in}} 
\caption{Surface brightness of optical lines in a simulation of a
photoionized nebula produced with {\it Cloudy} with a nebular hydrogen
density of 10~cm$^{-3}$. The solid line is for H$\beta$ $\lambda$4861,
the dashed line for He{\sc ii} $\lambda$4686, and the dotted line for
[O{\sc i}] $\lambda$6300.  The surface brightness was calculated by
integrating the emissivity predicted by {\it Cloudy} for a spherical
nebula with uniform density along lines of sight through the nebula. 
The actual nebula is not spherically symmetric and may have varying
density.} \label{cloudy} \end{figure}

\section{Interpretation}

We confirm the discovery by Pakull and Mironi (2002) of a He{\sc ii}
$\lambda 4686$ emission line nebula near the ultraluminous source in
Holmberg II.  Our improved X-ray astrometry derived from {\it Chandra}
strengthens the association of the nebula with the X-ray source.

We detect one point-like source within the body of the nebula.  This
source is likely the optical counterpart to the ULX.  For a distance
to  Holmberg II of 3.05~Mpc, the absolute magnitude $M_{V}$ of this
source is between $-5.5$ and $-5.9$ depending on whether there is
reddening within Holmberg II as discussed in section 2.3.  The absolute
magnitude and B-V color are consistent a range of spectral types from
O4V to B3 Ib.  We note that this spectral classification is valid only
if there is no significant contribution of optical light from the
accretion disk and if the star is not significantly affected by the
gravitational pull or X-ray emission from the compact object.
Alternatively, the optical emission may be reprocessed emission from an
X-ray illuminated accretion disk.  The ratio of observed X-ray to
optical flux and the B-V color are consistent with those found for
low-mass X-ray binaries in which the optical emission is dominated by
an X-ray illuminated disk \cite{vp95}.

Other optical counterpart searches have found early-type stellar
counterparts to ultraluminous X-ray sources.  The optical counterpart
to M81 X-6 \cite{fabbiano88} has colors consistent with an O8V star,
although there is some uncertainty in the spectral type due to
significant reddening \cite{liu02}.  For NGC 1313 X-2, only the R
magnitude of the counterpart has been measured; the absolute magnitude
is consistent with an early O main sequence star or an OB supergiant
\cite{zampieri04}. In NGC 5204, the optical counterpart found on
ground-based images \cite{roberts01} was resolved into two sources with
HST images \cite{goad02}.

For Holmberg II, Pakull and Mironi (2002) report that the He{\sc ii}
$\lambda 4686$ emission has a FWHM of $2.2\arcsec$ and that the [O{\sc
i}] is offset to the West of the He{\sc ii} emission.  Our HST images
confirm these basic results and provide much more detailed information
on the morphology of the nebula.

The maximum extent of the He{\sc ii} emission is to the West of the
central star and is about $1.7\arcsec$ or 26~pc for a distance to
Holmberg II of 3.05~Mpc.  The extent of the nebula towards the East is
shorter, about $1.0\arcsec$ or 15~pc.  To the West of the central star
(from NW to SW), the morphology of the H$\beta$ emission is similar to,
but more extended than, that of the He{\sc ii} emission and [O{\sc i}]
emission is low close to the star and stronger farther out,
particularly past the point where the He{\sc ii} emission peaks. This
is consistent with the behaviour expected for a photo-ionized nebula.
He{\sc ii} emission is produced in regions of high excitation and
should be concentrated near the central source. H$\beta$ is produced
over a broad range of excitation and should cover a larger spatial
range than He{\sc ii}. [O{\sc i}] is preferentially produced at lower
excitation than He{\sc ii} and should be produced in the outer regions
of the nebula.    Hence, the relative morphologies of the He{\sc ii},
H$\beta$, and [O{\sc i}] emission support the hypothesis that the
nebula is photoionized \cite{pakull02}.  Fig.~\ref{cloudy} shows the
surface brightness versus radius calculated for a photoionized nebula
(discussed further below) for comparison.

To the East and South of the central star, the morphology of the
H$\beta$ emission is very similar to that of the He{\sc ii} emission
and [O{\sc i}] emission is absent.  This is also consistent with a
X-ray photoionized nebula if the column density of the nebula
integrated along line of sight outward from the central star, is
sufficiently small that the entire nebula to the East and South is
excited to produce He{\sc ii} emission \cite{pakull02}.

We note that there is additional H$\beta$ emission to the NW of the
central star and beyond the extent of the He{\sc ii} emission which
appears to be part of a ring nebula surrounding a bright star visible
in the narrow V band image and unrelated to the He{\sc ii} nebula.
Also, the isolated He{\sc ii} bright spot to the SE of the central star
appears to be due to incomplete subtraction of a star visible in the 
narrow V band image.

If the nebula is powered by X-ray photoionization from the central
source, then the morphology of the nebula appears inconsistent with
narrow beaming of the X-ray emission.

The He{\sc ii} nebular emission is likely isotropic.  Therefore, the
measured flux should provide a good estimate of the true luminosity.  
For a distance to Holmberg II of 3.05~Mpc, the He{\sc ii} $\lambda
4686$ line flux found here for the nebula implies a line luminosity of
$2.7 \times 10^{36} \rm \, erg \, s^{-1}$.  This is consistent with the
value reported by Pakull \& Mironi (2002).

He{\sc ii} $\lambda 4686$ emission is produced via the recombination of
doubly ionized He$^{++}$ and acts as a photon counter of radiation
ionizing the nebula in the He$^{+}$ Lyman continuum above 54~eV.  
Following Pakull \& Mironi (2002), we use this property to estimate the
true luminosity of the X-ray source.  The major uncertainty in their
estimate was imprecise knowledge of the shape of the photon spectrum
illuminating the nebula.  Using the XMM-Newton spectra, we are able to
reduce this uncertainty.

We used the photoionization code {\it Cloudy} version 94.00
\cite{ferland01} to estimate the true X-ray luminosity based on the
measured He{\sc ii} luminosity.  We used a metallicity of $Z = 0.07
Z_{\sun}$ \cite{mirioni02}.  The relation between the total He{\sc ii}
flux and the total ionizing X-ray flux is not sensitive to the
metallicity.   We ran simulations over a range of hydrogen density
within the nebula from 1 to 100~cm$^{-3}$.  The relation between the
total He{\sc ii} flux and the total ionizing X-ray flux is not
sensitive to the density as long as the region of fully ionized He is
contained within the simulated nebula.  The spatial extent of the
He{\sc ii} emitting region is sensitive to the density.  A constant
density of 10~cm$^{-3}$ gave a reasonable match to the observed spatial
extent of the He{\sc ii} emission, and we use this value in the
simulations presented here.  In the modeling, we assumed a spherical
geometry with a filling factor of unity.  We assumed that there is no
absorption between the X-ray source and the nebula.  We also include
the photon flux of an O5V star modelled as a blackbody with a
temperature 42000~K and a luminosity of $3.2 \times 10^{39} \rm \, erg
\, s^{-1}$. Fig.~\ref{cloudy} shows the calculated surface brightness
versus apparent offset from the X-ray source.  The surface brightness
was calculated by integrating the emissivity along lines of sight
through a spherically symmetric nebula.

Table~\ref{specfits} shows the photoionization luminosities required to
produce the measured He{\sc ii} luminosity for the various best fit
Comptonization spectra.  The inferred photoionization luminosities
range from $3.7$ to $6.1 \times 10^{39} \rm \, erg \, s^{-1}$.  The low
luminosity value comes from the September 2002 observation during which
the source was in an unusual low/soft state \cite{dewangan04}.  The
correct X-ray spectrum to use in the photoionization modelling would be
a luminosity weighted average of observed spectra sampling a duration
comparable to the recombination time of He$^{++}$ in the nebula,
roughly 3000 years for an electron density of 10~cm$^{-3}$ and
temperature of 20,000~K.  The luminosity weighting would suggest a
photoionization luminosity near the high end of the range.  

To test the sensitivity of these results to the flux from the companion
star, we also made runs with the O5V star replaced by a B2 Ib star
modelled as a blackbody with a temperature 18,500~K and a luminosity of
$2 \times 10^{38} \rm \, erg \, s^{-1}$.  The inferred photoionization
luminosities increase by 12 to 16\% if a B2 Ib star is used in place of
the O5V star.  

As noted above, our photoionization model assumes a covering factor of
unity and sufficient depth radially so that the entire X-ray flux of
the central source is absorbed.  Inspection of the He{\sc ii} line
emission image shows that the covering factor is below unity.  Also, as
described above, [O{\sc i}] emission should be present at larger radii
than the He{\sc ii} emission if the nebula is not density bounded. 
Comparison of the  [O{\sc i}] and He{\sc ii} images shows the nebula is
density bounded to the East and South of the central star.  Therefore,
the photoionization luminosity quoted here should be considered a lower
bound to the total luminosity of the ionizing radiation.  Partial
absorption intrinsic to the X-ray source would harden the soft X-ray
flux illuminating the nebula and would increase the required
luminosity.  The unabsorbed X-ray luminosity in the Comptonization
models, assuming isotropic emission, ranges from 5 to $17 \times
10^{39} \rm \, erg \, s^{-1}$ for the three observations.  This is
consistent with the  luminosity inferred from the He{\sc ii} emission,
given that the He{\sc ii} emission likely underestimates the true
luminosity due to the covering factor of the nebula being less than one
and the nebula being density bounded along several lines of sight.

We conclude that the X-ray and He{\sc ii} data are consistent with the
X-ray source being the ionization source powering the line emission
from the nebula.  The luminosity in He{\sc ii} emission is more than
an  order of magnitude greater than that observed from a similar nebula
around the black hole candidate LMC X-1 \cite{pakull86}, suggesting
that Holmberg II contains a much more powerful X-ray source.  We
estimate that the total luminosity of the X-ray source is at least $3.3
\times 10^{39} \rm \, erg \, s^{-1}$ and likely above $6 \times 10^{39}
\rm \, erg \, s^{-1}$ if the extrapolation of the X-ray spectrum
between 54~eV and 300~eV is accurate.  This suggests that the compact
object is truly ultraluminous.  Assuming an Eddington luminosity of
$1.3 \times 10^{38} \rm \, erg \, s^{-1} M_{\odot}^{-1}$, the compact
object mass would be at least $25 M_{\odot}$ and likely above $40
M_{\odot}$.  This may suggest a compact object which is more massive
than the stellar-mass black holes candidates in our Galaxy, for which
all of the dynamically measured masses lie below $20 M_{\odot}$
\cite{mcclintock03}.

\section*{Acknowledgments}

We thank the Aspen Center for Physics for its hospitality during the
workshop ``Compact Objects in External Galaxies'' and an anonymous
referee for comments which improved this manuscript.  PK thanks Manfred
Pakull for useful discussions and Manfred Pakull and Laurent Mirioni
for a copy of Laurent Mirioni's thesis.  PK acknowledges partial
support from STScI grant HST-GO-09684.01 and Chandra grant CXC
GO3-4052X.  AZ acknowledges partial support from NASA LTSA grant
NAG5-13056.  STSDAS and PyRAF are products of the Space Telescope
Science Institute, which is operated by AURA for NASA.


\label{lastpage}

\end{document}